\documentclass[%
pra,
 amsmath,amssymb,
 reprint,showkeys, %
superscriptaddress,longbibliography
]{revtex4-2}
\usepackage{braket}
\usepackage{hyperref}
\usepackage{ulem}
\usepackage{mathtools}
\usepackage{graphicx} % Include figure files
\usepackage{epsfig}
\usepackage{epstopdf}
\usepackage{dcolumn}% Align table columns on decimal point
\usepackage[dvipsnames]{xcolor}
\usepackage{bm}% bold math
\graphicspath{{fig/}}

\usepackage{appendix}

\begin{document}

\preprint{AIP/123-QED}

\title[]{ Long persistent anticorrelations in few-qubit arrays
}% Force line breaks with \\
%\thanks{To whomever}

\author{Danil Kornovan}%
\affiliation{School of Physics and Engineering, ITMO University, Lomonosova st. 9, Saint-Petersburg, Russia, 197101}
\email{d.kornovan@metalab.ifmo.ru}
%\affiliation{\affilITMO %\\This line break forced with \textbackslash\textbackslash
% }%

\author{Alexander Poddubny}
\affiliation{Department of Physics of Complex Systems, Weizmann Institute of Science, Rehovot 7610001, Israel}

%\affiliation{\affilNone}
% \affiliation{\affilIoffe%\\This line break forced% with \\
% }%
% \affiliation{\affilITMO %\\This line break forced with \textbackslash\textbackslash
% }%
\author{Alexander Poshakinskiy}
 \affiliation{Ioffe Institute, 26 Politekhnicheskaya st., 194021 St.-Petersburg, Russia}%Lines break automatically or can be forced with \\

%\date{}% It is always \today, today,

\begin{abstract}
We consider theoretically the mechanisms to realize antibunching between the photons scattered on the array of two-level atoms in a general electromagnetic environment. Our goal is the antibunching  that persists for the times much longer than the spontaneous emission lifetime of an individual atom.  We identify two  mechanisms for such persistent antibunching. The first one is based on subradiant states of the atomic array, and the second one does not require any subradiant states. We provided two specific  examples of array parameters with optimized antibunching, based on  an array in a free space and an array coupled to a waveguide.
 % We analyze the explicit analytical expression for a two-photon scattering amplitude on an ensemble of two-level atoms, and based on this analysis, we suggest two mechanisms to achieve photon antibunching that is persistent on large delay times. The first mechanism is based on the usage of single-excitation collective subradiant states of the system, when only a single long-lifetime eigenstate contributes significantly to the scattering amplitude. The second mechanism relies on a destructive interference between the contributions of at least several collective states, none of which are necessarily strongly subradiant. For both cases we provide pictorial examples of how to achieve such an antibunching based on interacting two-level atoms in free-space, and in waveguide quantum-electrodynamical setup in a reflection geometry.
\end{abstract}

%Use showkeys class option if keyword
\maketitle

 \section{Introduction.} 
Last several years have seen a  revival of interest to the quantum optics of arrays of natural and artificial atoms~\cite{Rui2020}. This has been stimulated by the emergence of highly coherent and ordered structures, such as  lattices of trapped atoms in free space \cite{BarredoSci2016, BarredoNat2018, SchlosserJOPB2020, SchlosserPRL2023}, or emitters coupled to a waveguide~\cite{GonzalesNatPhot2015,Roy2017,Tiranov2023,Kannan2023,sheremet2021waveguide}. One of the opportunities, offered by ordered atomic structures for quantum optics, is the on-demand generation of quantum  states of light, including squeezed light~\cite{SolomonsArxiv2021}, cluster states~\cite{BekensteinNatPhys2020,Guimond2020,Zeiher2022} and antibunched photons~\cite{prasad2020correlating}.

While the photon  antibunching, i.e. suppressed photon-photon correlation function $g^{(2)}(\tau)$ for zero delay between the photons $\tau=0$, is probably the simplest and the most known example of quantum photon correlations, even  the antibunching  optimization  problem  is not trivial, and there is an extensive amount of work being done and is ongoing in this direction \cite{ShenPRL2007, NayakPRAR2009, ZhengPRA2012, Fang2014, FangPRA2015, CidrimPRL2020, WilliamsonPRl2020, PedersenArxiv2022, ZhangQuantum2022, VlasiukArxiv2023}. Indeed, for practical applications it is desirable to realize the antibunching that is  not only strong, $g^{(2)}(0)\ll 1$, but also persistent, i.e. $g^{(2)}(\tau)\ll 1$ for a considerable range of delays between the photons $\tau$. To the best of our knowledge, there is no general universal scheme how to realize such persistent antibunching. Here, we discuss two specific mechanisms for persistent antibunching in atomic arrays. The first mechanism is relatively straightforward and based on formation of subradiant states in the array, where the spontaneous decay is suppressed by destructive interference~\cite{Brewer1996,Kaiser2016}. Indeed, it is  expected that a resonant excitation of subradiant states can provide long-living quantum correlations, including antibunching, see e.g. Ref.~\cite{Facchinetti2016,Asenjo2017,WilliamsonPRl2020,Poshakinskiy2021Borrmann,Poddubny2022}. The second mechanism for persistent antibunching is somewhat more surprising. Namely, we show that it is possible to achieve $g^{(2)}(\tau)\ll 1$ for the times $\tau$ that are noticeably longer than the lifetimes of all the individual eigenstates of the system. In another words, such method does not require any subradiant states to exist. Instead, the antibunching relies on the peculiar destructive interference between single-excited states with different lifetimes, so that their contributions to the photon-photon correlations suppress each other in a considerable time range, even though neither of them is required to be subradiant.
While we have performed the optimization for specific cases of an array in a free space  and an array, coupled to a waveguide, the proposed approaches are rather general and are not inherently  restricted to particular geometries. Thus, we hope that our results could provide useful insights for design of future quantum devices.

\begin{figure}[tb]
\includegraphics[width=0.4\textwidth]{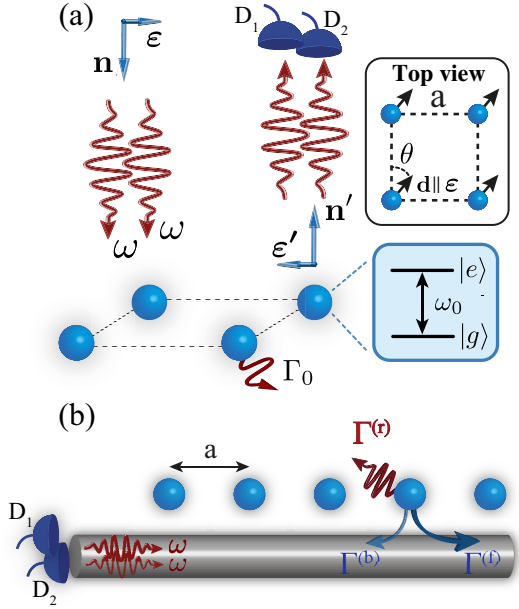}
\caption{Two schemes considered in this work. (a) A four-qubit system in free-space arranged in a rectangle with nearest-neighbor distance $a$. The two  photons of frequencies $\omega$ are incident normally along the direction $\bm n$ , while the outgoing photons are measured by two detectors $D_{1}, D_{2}$. The incident photon polarization $\bm{\varepsilon}$ and atomic dipole moments $\bm d$ are parallel to each other and their orientation is characterized by an angle $\theta$. (b) One-dimensional periodic array of qubits, asymmetrically coupled to a single guided mode with forward (backward emission rates) $\Gamma^{(\text{f})}(\Gamma^{(\text{b})})$, while the emission rate out of a waveguide is $\Gamma^{(\text{r})}$. The scattered photons are measured in the reflection geometry. }
\label{F1}
\end{figure}
 
 The rest of the manuscript is organized as follows. We start in Sec.~\ref{sec:theory} by outlining a  theoretical framework to calculate the photon-photon scattering matrix 
 (Sec.~\ref{sec:theory1}) and photon-photon correlations  (Sec.~\ref{sec:theory2}).  Next, we present in Sec.~\ref{sec:results} the results for optimized persistent antibunching in two different geometries. Section~\ref{sec:results1} considers a four-atom array in free space [Fig.~\ref{F1}(a)] and Sec.~\ref{sec:results2} is devoted to atoms coupled to photons in a waveguide [Fig.~\ref{F1}(b)]. The results are summarized in Sec.~\ref{sec:conclusions}.
 The auxiliary theoretical details are given in Appendices~\ref{App A} and \ref{App B}.

\section{Theoretical framework}\label{sec:theory} 
\subsection{General formulas for two-photon scattering in frequency domain}\label{sec:theory1}
A Green function approach to describe for the two-photon scattering on an arbitrary array of atoms coupled to the waveguide has been developed in Refs.~\cite{PoshakinskiyPRA2016, ke2019inelastic,Fang2014,Laakso2014}.  The photon Green function, defined by the vector equation
\begin{multline}
    \nabla \times \nabla \times \mathbf{G}(\mathbf{r}, \mathbf{r}', \omega) - \dfrac{\omega^2}{c^2} \bm{\varepsilon}(\mathbf{r}, \omega) \mathbf{G}(\mathbf{r}, \mathbf{r}', \omega)  \\=
    \mathbf{I} \delta(\mathbf{r} - \mathbf{r}'),
\end{multline}
can describe a linear response in an arbitrary electromagnetic environment characterized by the dielectric permittivity tensor $\bm{\varepsilon}(\mathbf{r},\omega)$. Thus, it is straightforward to generalize the photon scattering calculation from the waveguide to a more general setup. In this section we consider two-photon scattering for an array of $N$ atoms located at points $\mathbf{r}_i$ ($i=1\ldots N$) in free space, as shown in Fig.~\ref{F1} (a).

The wave function of the scattered photon pair $\Psi$ is related to that of the incident pair $\Psi^{(0)}$ via the two-photon scattering matrix:
\begin{equation}\label{eq:Psi}
    \Psi_{\lambda_1',\lambda_2'} = \frac12 \sum_{\lambda_1\lambda_2} S_{\lambda'_1, \lambda'_2; \lambda_1, \lambda_2} \Psi^{(0)}_{\lambda_1\lambda_2},
\end{equation}
where $\lambda$ is the multi-index incorporating all of the quantum numbers of the photonic eigenmodes, i.e., for the photons in free space $\lambda =(\bm k, \sigma)$, $\bm k = k \bm n$ is the wave vector, $\bm n$ is the propagation direction, $\sigma$ is the polarization index, $k= \omega_{\lambda}/c$, $\omega_{\lambda}$ is the mode frequency.
% The scattering is characterized by the two-photon $S$ matrix
% defined as 
% \begin{equation}
%     \text{\red {please add definition with correct factors $2\pi$ etc. Maybe Sasha V. could help here}}
% \end{equation}

The $S$-matrix can be represented as a sum of the coherent  independent photon scattering term, and the term responsible for an inelastic scattering and photon-photon interaction: 
\begin{align}\label{Sdefs}
    &S_{\lambda'_1, \lambda'_2; \lambda_1, \lambda_2} = S^{(\text{lin})}_{\lambda'_1, \lambda'_2; \lambda_1, \lambda_2} + S^{(\text{nlin})}_{\lambda'_1, \lambda'_2; \lambda_1, \lambda_2}, \\
    &S^{(\text{lin})}_{\lambda'_1, \lambda'_2; \lambda_1, \lambda_2} =  \left( S^{(1)}_{\lambda'_1, \lambda_1} S^{(1)}_{\lambda'_2, \lambda_2} + S^{(1)}_{\lambda'_1, \lambda_2} S^{(1)}_{\lambda'_2, \lambda_1} \right), \nonumber\\
    &S^{(\text{nlin})}_{\lambda'_1, \lambda'_2; \lambda_1, \lambda_2} = 4 \pi \delta(\Omega' - \Omega)\nonumber\\ \nonumber
    & \times\sum_{i,j=1}^N
    s^{+}_{\lambda'_1, i}(\omega_{\lambda_1'}) s^{+}_{\lambda'_2, i}(\omega_{\lambda_2'}) Q_{i,j}(\Omega) s^{-}_{\lambda_1, j}(\omega_{\lambda_1}) s^{-}_{\lambda_2, j}(\omega_{\lambda_2}),
\end{align}
where 
%$\lambda$ is the multi-index incorporating all the quantum numbers of the photons,
$S^{(1)}_{\lambda', \lambda}$ is a single photon $S$-matrix, $Q_{i,j}(\Omega)$ is the two-photon scattering kernel matrix, indices $i,j=1\ldots N$ enumerate atoms, $\Omega = \omega_{\lambda_1} + \omega_{\lambda_2}$, $\Omega' = \omega_{\lambda_1'} + \omega_{\lambda_2'}$, $s^{-}_{\lambda, i}$  represents the self-consistent excitation amplitude of atom $j$ due to the absorption of a photon $\lambda$, and $s^{+}_{\lambda, i}$ describes the amplitude of the inverse process.

The amplitudes $s^{\pm}_{\lambda, i}$ can be calculated as:
\begin{align}
    s^-_{\lambda, i}(\omega_\lambda) &= \sum\limits_{j=1}^{N} G_{ij}(\omega_{\lambda}) \ \mathbf{d}_j \cdot \mathbf{E}_{\lambda}(\mathbf{r_j}),\nonumber\\
    s^+_{\lambda, i}(\omega_\lambda) &= \sum\limits_{j=1}^{N} 
   \mathbf{E}_{\lambda}^*(\mathbf{r_j}) \cdot \mathbf{d}_j^*
   \  G_{ji}(\omega_{\lambda}),
    \label{spsmdefs}
\end{align}
where
$\mathbf{E}_{\lambda}$ is the electric field amplitude of mode $\lambda$, $\mathbf{d}_j$ the matrix element of the dipole operator between the ground and excited state of atom $j$, and  $G_{ij}(\omega)$ is the 
$N \times N$ matrix of the single-excitation quantum-mechanical Green's function, that is readily calculated as $G(\omega) = \left(  \omega -  H^{\text{(eff)}} \right)^{-1}$, where the effective  Hamiltonian accounts for the transfer of excitation between the atoms via the electromagnetic field and reads~\cite{ShahmoonPRA2013, GonzalesNatPhot2015, KornovanPRB2016}
\begin{align}\label{eq:Heff}
     H^{(\text{eff})}_{ij}  =   
     \begin{cases}
    -\frac{4 \pi \omega_0^2}{c^2} \mathbf{d}^{\dagger}_i \mathbf{G}(\mathbf{r}_i, \mathbf{r}_j, \omega_0) \mathbf{d}_j, & \text{ if $i \ne j$}, \\
     \omega_0-i  \frac{\Gamma}2, & \text{ if $i = j$},
    \end{cases}
\end{align}
$\omega_0$ is the atomic transition frequency and $\Gamma$ is the spontaneous decay rate of the atomic excited state. We do not restrict ourselves to the reciprocal systems, which means that the matrix $H_{i,j}^{(\text{eff})}$ and $G_{ij}(\omega)$ are of general form, neither Hermitian nor symmetric. Moreover, in the definition of $G(\omega)$ above we omitted $\hbar$ as we regard it being equal to unity throughout the manuscript. 

%Specific form of this constant depends on the setup, and will be given later.

The $N \times N$ scattering kernel matrix $Q_{ij}(\Omega)$ 
for a collection of two-level atoms
can be expressed via the single-excitation Green's function as~\cite{PoshakinskiyPRA2016,ke2019inelastic}:
\begin{align}
    &Q(\Omega) = \Sigma^{-1}(\Omega), \nonumber\\ &\Sigma_{ij}(\Omega) = \int G_{ij}(\omega) G_{ij}(\Omega - \omega) \frac{d \omega}{2 \pi}.
    \label{Qdef}
\end{align}
%where $G(\omega) = \left( \omega - H^{(\text{eff})} \right)^{-1}$ is the single-excitation Green's operator, $\odot$ is a Hadamard (element-wise) product, and $H^{(\text{eff})}$ is the matrix of the effective Hamiltonian on a single excitation domain, which can be expressed solely through the electrodynamical Green's tensor\cite{ShahmoonPRA2013, GonzalesNatPhot2015, KornovanPRB2016}:
% \begin{gather}\label{eq:Heff}
%     \langle e_i | H^{(\text{eff})} | e_j \rangle = \begin{cases}
%     - 4 \pi k_0^2 \mathbf{d}^{\dagger}_i \mathbf{G}(\mathbf{r}_i, \mathbf{r}_j, \omega_0) \mathbf{d}_j, & \text{ if $i \ne j$}, \\
%     - i \hbar \dfrac{\Gamma_j}{2}, & \text{ if $i = j$}
%     \end{cases},
% \end{gather}
%where  $k_0 = \omega_0/c$ is the resonant wavenumber, $\mathbf{d}_j$ is the dipole moment of atom $j$, and $\mathbf{r}_j$ is its location, while $\Gamma_j$ is the total spontaneous emission rate of that atom.
%However, from now on we will assume that all atoms have the  same emission rate $\Gamma_j = \Gamma$ for simplicity.
%
In cases when the effective Hamiltonian matrix $H_{i,j}^{(\text{eff})}$ is diagonalizable one can find $\Sigma(\Omega)$ as the expansion over its right ($v_i^{(R, \nu)}$) and left ($v_i^{(L, \nu)}$) eigenvectors (enumerated by $\nu=1\ldots N$), by using the eigenexpansion of the single excitation Green's function $G_{i,j}(\omega') = \sum\limits_{\nu'=1}^{N} \dfrac{v^{(R,\nu')}_{i} v^{(L,\nu')}_{j}}{\omega' - E^{(1,\nu')}} =  \sum\limits_{\nu'=1}^{N} \dfrac{g^{(\nu')}_{i,j}}{\omega' - E^{(1,\nu')}}$, and taking the frequency integral by the residue theorem, one can write:
\begin{gather}
\Sigma_{ij}(\Omega) = - i \sum\limits_{\nu_1, \nu_2} \dfrac{ g^{(\nu_1)}_{i,j} g^{(\nu_2)}_{i,j} }{\Omega - E^{(1, \nu_1)} - E^{(1, \nu_2)}},
    % \Sigma_{ij}(\Omega) = - i \sum\limits_{\nu_1, \nu_2} \dfrac{v^{(R, \nu_1)}_{i} v^{(L, \nu_1)}_{j} v^{(R, \nu_2)}_{i} v^{(L, \nu_2)}_{j} }{\Omega - E^{(1, \nu_1)} - E^{(1, \nu_2)}},
    \label{Sig1exc}
\end{gather}
where $E^{(1, \nu)}$ are the corresponding complex eigenvalues.
%where $\mathbf{v}^{(R, \nu)} (\mathbf{v}^{(L, \nu)})$ is the right (left) single excitation eigenstate, with the corresponding complex valued eigenegergy $E^{(1, \nu)}$.
We want to note that there is a way to define $Q(\Omega)$ that is alternative to Eq.\:\eqref{Qdef} which includes explicitly the two-excitation eigenstates of the problem, and for that we address the reader to App. \ref{App B}.

Now we have all the ingredients to discuss two-photon detection, which will be covered in the following subsection.
%%%%%%%%%%%%%%%%%%%%%%%%%%%%%%%%%%%%%%%%%%%%%%%%
\subsection{Two-photon detection in the time-dependent correlations}\label{sec:theory2}
%%%%%%%%%%%%%%%%%%%%%%%%%%%%%%%%%%%%%%%%%%%
In this work, we consider a continuous-wave excitation where the incident  photons are infinitely delocalized along the direction of propagation and described by well-defined quantum numbers $\lambda_{1,2}$. The quantum correlations for the scattered light are then characterized by the probability to detect two photons at times $t_1$ and $t_2$, that depends only on the time difference $t_1-t_2$. To calculate this probability,  we perform a double Fourier transform of the scattered-photon wave function Eq.~\eqref{eq:Psi} and obtain the two-photon wave function in the time domain:
\begin{multline}
     \Psi(t_1, t_2)  = \\
    \frac{1}{ 8 \pi^2}  \iint S_{\lambda'_1, \lambda'_2;\lambda_1, \lambda_2} e^{-i (\omega_{\lambda'_1} t_1 + \omega_{\lambda'_2} t_2)} d \omega_{\lambda_1'} d \omega_{\lambda_2'} ,
    \label{Psit1t2}
\end{multline}
where the integration shall be carried over the frequencies corresponding to modes with the fixed propagation directions $\bm n_1'$ and $\bm n_2'$. The normalized second-order correlation function is given by $g^{(2)}(t_1, t_2)  = \left| \Psi(t_1, t_2)/\Psi^{(\text{lin})}(t_1,t_2) \right|^2$, where $\Psi^{(\text{lin})}(t_1,t_2)$ is obtained from Eq.~\eqref{Psit1t2} by replacing $S$ with  $S^{(\text{lin})}$. 

For the purpuses of this work, we will consider the case when both incident photons, and detected photons are pairwise identical in their quantum numbers: $\lambda_1 = \lambda_2 = \lambda$, $\lambda'_1 = \lambda'_2 = \lambda'$. In this case the $g^{(2)}(t_1, t_2)$ function can be presented in the following form (see Appendix \ref{App A} for the derivation):
\begin{gather}
g^{(2)}_{\lambda';\lambda}(\tau) = \left| 1 - \sum\limits_{\nu} C^{(\nu)}_{\lambda';\lambda} e^{-i \left( E^{(1, \nu)} - \omega \right) \tau} \right|^2,
    \label{g2simple}
\end{gather}
where $C^{(\nu)}$ are the constants determined by the residues of the integrand in \eqref{Psit1t2} at the resonances of singly-excited states.  By a simple observation, one can see that in order to realize antibunching at zero delay $g^{(2)}(0)\ll 1$, one needs to achieve $\left| 1 - \sum_{\nu} C^{(\nu)}_{\lambda'_1,\lambda'_2;\lambda_1,\lambda_2} \right| \ll 1$. It is not immediately obvious how to satisfy such an inequality because the complex-valued amplitudes $C^{(\nu)}$ in general case have a rather intricate dependence on single-, and two-excitation eigenstates, as well as on the scattering setup, that is  propagation directions and polarizations of the incoming and outgoing detected photons. The complexity of the explicit form of $C^{(\nu)}_{\lambda'_1,\lambda'_2;\lambda_1,\lambda_2}$ makes it hard to analyze for a general setup, however, for certain few-atom systems the analysis can be insightful, as we will show later.

% In order to make the antibunching  persistent  at long times $ \tau > 1/\Gamma$ ($\Gamma$ is the emission rate of an isolated qubit), one needs to be in resonance with  subradiant states, that is to satisfy the condition $\left|\omega - \text{Re}\: E^{(1, \nu)}\right| \ll \Gamma$, while simultaneously $\text{Im}\: E^{(1, \nu)} \ll \Gamma$. We stress that these two conditions must be simultaneously satisfied for all single excitation states $\nu$ that have a significant contribution to the detected signal $|C^{(\nu)}_{\lambda'_1,\lambda'_2;\lambda_1,\lambda_2}| \sim 1$. Designing an ensemble of qubits along with the scattering geometry in such a way appears to be non-trivial. While we do not aim to solve this problem in general case, we suggest one of straightforward solutions. One needs to setup the system in such way that there is a distinct subradiant single excitation state that is spectrally well isolated in the single excitation domain, and provides the main contribution most to the signal. In this case one only needs to tune the frequency of both incident photons to the frequency of this subradiant state $\omega = \text{Re}\: E^{(1, \nu)}$. The  only remaining required condition is then the presence of strong anti-correlations  at zero delay $g^{(2)}(0) \approx 0$. In the next section we will provide a two conceptual example of such a procedure for two different systems.
In order to make the antibunching  persistent  at long times $ \tau > 1/\Gamma_0$ ($\Gamma_0$ is the emission rate of an isolated qubit), one needs to be in resonance with  subradiant states, that is to satisfy the condition $\left|\omega - \text{Re}\: E^{(1, \nu)}\right| \ll \Gamma$, while simultaneously $-2 \text{Im}\: E^{(1, \nu)} \ll \Gamma$. We stress that these two conditions must be simultaneously satisfied for all single excitation states $\nu$ that have a significant contribution to the detected signal $|C^{(\nu)}_{\lambda'_1,\lambda'_2;\lambda_1,\lambda_2}| \sim 1$. Designing an ensemble of qubits along with the scattering geometry in such a way appears to be a non-trivial problem in general. While we do not aim to explore all possibilities, we suggest a straightforward solution to this problem. One can setup the system in such a way that there is a distinct subradiant single excitation state that is spectrally well isolated in the single excitation domain, and provides the main contribution to the signal~\cite{WilliamsonPRl2020}. In this case one only needs to tune the frequency of both incident photons to the frequency of this subradiant state $\omega = \text{Re}\: E^{(1, \nu)}$. The  only remaining required condition is then the presence of strong anti-correlations  at zero delay $g^{(2)}(0) \approx 0$. In the next section we will provide a simple conceptual example of such a procedure.

\section{Results and discussions}\label{sec:results}

\subsection{Example 1: Four atoms in free space}\label{sec:results1}

% \begin{figure}[t]
% \includegraphics[width=0.48\textwidth]{F22.eps}
% \caption{(a) Scheme of the system. (b) Second-order correlation function $g^{(2)}$ over delay time $\tau$, the inset shows the oscillating behavior close to zero delay $\tau$. (c) Single exciation spectrum of the effective Hamiltonian (eigenfrequencies $\Delta\omega^{(1,\nu)}$, and emission rates $\Gamma^{(1, \nu)}$), the color encodes the contribution to the $g^{(2)}$ signal that is a constant $c^{(1, \nu)}$ given in Eq. \eqref{g2simple}. (d) Normalized two-photon scattering amplitudes: single-atom non-linearity (1 at. nlin.), contribution from doubly-excited state of $H_{\text{eff}}$ (2 ex. st.), and the total amplitude (the quantity being taken $|\dots|^2$ in Eq. \eqref{g2simple}) at delay time $\tau = 0$. {\color{red} parameters} }
% \label{F2}
% \end{figure}

\begin{figure}[t]
\includegraphics[width=0.48\textwidth]{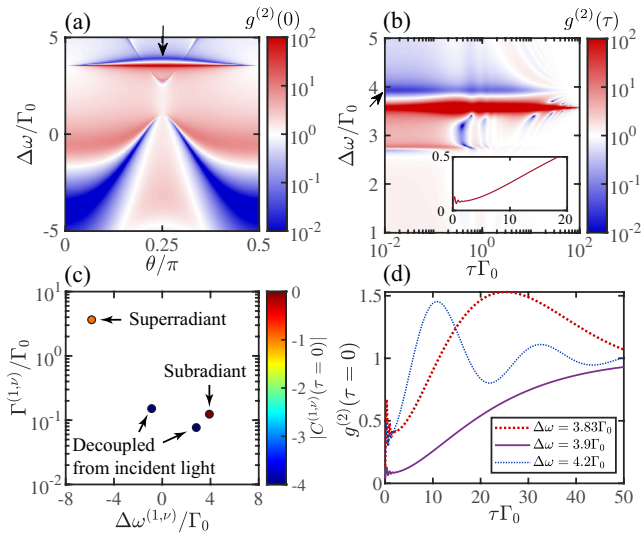}
\caption{(a) $g^{(2)}(\tau=0)$ dependence for free-space arrangement as a function of detuning of photons $\Delta\omega$, and polarization angle $\theta$. Here the distance between neighboring atoms is $a=0.1\lambda_0$. (b) Second-order correlation function $g^{(2)}$ versus the detuning of both photons $\Delta\omega$ and delay time $\tau$, the inset shows the behavior of $g^{(2)}(\tau)$ for $\Delta\omega=3.9\Gamma_0$ specified by an arrow.
(c) Single excitation spectrum of the effective Hamiltonian (eigenfrequencies $\Delta\omega^{(1,\nu)}=\text{Re\,}E^{(1, \nu)}-\omega_0$, and emission rates $\Gamma^{(1, \nu)}=-2\text{Im\,}E^{(1, \nu)}$), the color encodes the contribution to the $g^{(2)}$ signal that is a constant $C^{(\nu)}$ given in Eq. \eqref{g2simple}.
(d) Temporal dynamics of $g^{(2)}(\tau)$ for different detunings: $\Delta\omega=3.83\Gamma_0$ (red thick dotted), $\Delta\omega=3.9\Gamma_0$ (purple thick solid), $\Delta\omega=4.2\Gamma_0$ (blue thin dotted).  }
\label{F2}
\end{figure}

Our first example is based on four two-level atoms that are arranged in a square in free space, as shown in Fig.~\ref{F1}(a). The atoms are at a strongly subwavelength distance between the nearest neighbors $a = 0.1 \lambda_0 \equiv 0.1\times 2\pi c/\omega_0$. This  condition is required in order for subradiant states to arise in the system. We assume that the transition dipole moments of the atoms are linearly polarized in the plane and characterized by orientation angle $\theta$. The photons are incident normally on the system ($\mathbf{n}_1 = \mathbf{n}_2 = \mathbf{n} || \mathbf{e}_z$) being polarized parallel to the dipole moments. We detect the back-scattered photons, $\mathbf{n}_1'=\mathbf{n}_2' = - \mathbf{n}$, with the same polarization as the incident ones.
%\sout{In case of free space plane waves the photon mode is characterized by the polarization $(\text{index}~\sigma,  \text{ polarization vector }  \bm{\varepsilon}$), frequency ($\omega$), and the direction of propagation $\mathbf{n}$. In this work we consider a simple case when in each pair (incident, and outgoing) photons are identical, which means they have the same frequency, polarization and propagation direction. Moreover, here we will study the reflection geometry, when both outgoing photons are propagating in the $\mathbf{n'} = - \mathbf{n}$ direction.}

We begin with examining the equal-time photon-photon correlations dependence on the incident photon frequency and on their polarization, see Fig.~\ref{F2}(a). The results demonstrate that by tuning the polarization angle $\theta$ and detuning from resonance $\omega_{1}-\omega_0=\omega_{2}-\omega_0 \equiv \Delta\omega$, it is possible to achieve $g^{(2)}(0)<1$ with different sets of parameters. The parameters we want to focus on  are denoted by a black arrow for $\theta = 0.25 \pi$, $\Delta\omega = 3.9\Gamma_0$. We can also observe the map of $g^{(2)}(\tau)$ displayed in Fig. \ref{F2} (b) versus both detunings of photons $\Delta\omega$ and time delay $\tau$. One can see that, in general, second-order correlation function can demonstrate a quite complicated dynamics that switches between strong bunching and antibunching. What is especially interesting is the dynamics for the aforementioned detuning value $\Delta\omega = 3.9 \Gamma_0$ (black arrow), when the scattered photons remain antibunched for a long time. Indeed, as the inset demonstrates, the $g^{(2)}$ function exhibits a very slow growth towards unity, achieving the value $0.5$ only at the time approximately equal to $\tau \approx 18/\Gamma_0$. On top of this persistent antibunching there are some oscillations with a small amplitude for $\tau \Gamma_0 < 2$. This behavior can be clarified with the help of Fig.~\ref{F2} (c), where the single excitation spectrum of $H^{(\text{eff})}$ is presented. As can be seen, out of  four collective eigenstates in total, only two have a non-zero contribution to the signal. The other two are antisymmetrically excited dimers consisting of atoms on the two diagonals -- such states  can not be excited with the perpendicularly incident plane wave because of the selection rules.  One of the states that are coupled to light is a subradiant state with $\Delta\omega^{(1,-)} \approx 0.39 \Gamma_0$, $\Gamma^{(1, -)} \approx 0.12 \Gamma_0$. There is also a non-zero overlap with the quasi-superradiant (bright) state that is strongly shifted in frequency: $\Delta\omega^{(1,+)} \approx -5.85 \Gamma_0$, $\Gamma^{(1,+)} \approx 3.64 \Gamma_0$. The corresponding $C^{(\nu)}_{\lambda',\lambda}$ constants for these states are given by $C^{(-)}_{\lambda',\lambda} \approx 1.07 e^{i 0.91 \pi}$, $C^{(+)}_{\lambda',\lambda} \approx 0.115 e^{-i 0.485 \pi}$, from which one can immediately conclude that we are predominantly exciting the subradiant state as $|C^{(-)}_{\lambda',\lambda} | \gg |C^{(+)}_{\lambda',\lambda} |$, and the quasi-superradiant state gives a rather small contribution giving rise to fast oscillations seen on the inset of Fig.~\ref{F2}(b) owing to a large detuning $\Delta\omega^{(1,-)}-\Delta\omega^{(1,+)} \gg \Gamma_0$. After the contribution of the superradiant state decays out due to its small lifetime, the excitation resides in the remaining subradiant state, and its long lifetime provides long-term anticorrelations. Finally, we also present in Fig. \ref{F2} (d) temporal dynamics of $g^{(2)}(\tau)$ function for detunings around the so far discussed value $\Delta\omega = 3.9 \Gamma_0$ close to the subradiant state. For the reasons already discussed before, as one detunes from the frequency of the subradiant state, temporal oscillations appear, as well as the zero-delay value of $g^{(2)}(0)$ is increased. The latter happens as by moving off the resonance with the state, the $C^{(\nu)}_{\lambda',\lambda}$ constants are altered (Eqs. \ref{g2simple} and \ref{Cconst}), making the destructive interference between the linear and non-linear signals less prominent. Nonetheless, if the detuning from the subradiant state is not too large, it is still possible to see anticorrelations $g^{(2)}(\tau)<1$ that survive over a few atomic lifetimes $\tau_0 = 1/\Gamma_0$. The width of this frequency window in which anticorrelations are observed can be very roughly approximated by the linewidth of the subradiant state being excited, however, it also depends on where spectrally other states contributing to the signal are. For instance, as there are no other states to the blue region of the subradiant state, anticorrelations persist over larger blue detunings (up to $\Delta\omega = 4.2 \Gamma_0$) rather than red ones (only around $\Delta\omega = 3.83 \Gamma_0$).

The demonstrated subradiance-induced antibunching in  a small array of atoms in a free-space has a lot of similarities with the one recently studied in  Ref.~\cite{WilliamsonPRl2020}. There are however two important differences. First, our theoretical approach allows us to rigorously obtain an explicit expansion of the correlation function over singly- and doubly-excited eigenstates. In our opinion this is a  more suitable technique  to identify the role of various eigenstates in the correlation than the semi-phenomenological expansion  in Ref.~\cite{WilliamsonPRl2020}. This could provide further insight into the ongoing experimental investigations  of the  quantum  light scattering from the  regular free-space arrays~\cite{Rui2020,Zeiher2022}. Second, here we analyze the dependence of the correlations on the azimuthal polarization angle of incoming photons, while Ref.~\cite{WilliamsonPRl2020} has been focused on the specific case $\theta=0$. As shown in Fig.~\ref{F2}b, tuning the polarization angle to the value of $\theta=\pi/4$ significantly enhances the antibunching.

We would also like to discuss the linewidth of the excitation source that 
is required in order to observe the  antibunching. Since the mechanism considered in this section relies on a resonantly excited subradiant state,  it is reasonable to assume that the precision of the frequency tuning has to be approximately equal to the emission rate of the corresponding state.   The linewidth of the optical transition for Cs or Rb atoms is about ~$5\div6$~MHz~\cite{sheremet2021waveguide}.  Even comercially available optical lasers allow for the stabilization on the order of ${1\div10}$~kHz which is smaller by 3 orders of magnitude than the spontaneous emission linewidth, while modern set-ups demonstrate Hz or even sub-Hz stabilization \cite{LiangOptLett2023}. For superconducting qubits the transition linewidth for the emission into the waveguide is  typically ${2\div 3}$ times larger~\cite{sheremet2021waveguide}, while the microwave driving sources are  much more frequency stable with the  sub-Hz frequency fluctuations.

So far we have discussed the case of free space four-qubit ensemble, where the long persistent antibunching of photons appears due to a single excitation subradiant state that controls the scattering, while all other states have way smaller contributions. Even though this suggestion came naturally from the form of Eq. \eqref{g2simple}, it turns out to be not the only way to achieve the desired long temporal anticorrelations.
 % {\color{red} We must note that the results of this section have a significant overlap with the numerical studies done in Ref. \cite{WilliamsonPRl2020} on the ''superatom`` picture in ensembles under a low intensity light illumination, when the whole atomic subsystem is presented as a single emitter with altered frequency and linewidth, giving rise to short-lived or long-lived antibunching depending on whether the excited state is super- or subradiant. In this work, however, we present a rigorous, ab-initio proof of this statement for a two-photon scattering.}

In the next subsection we will have a look at another system based on waveguide setup, where the mechanism is different.

\subsection{Example 2:  Asymmetric waveguide-QED}\label{sec:results2}
%\commentANP{I've removed ``accidental destructive interference'' from the title  since the whole paper is kind of about it}
%%%%%%%%%%%%%%%%%%%%%%%%%%%%%%%%%%%%%%%%%%%%%
\begin{figure}[t]
\includegraphics[width=0.48\textwidth]{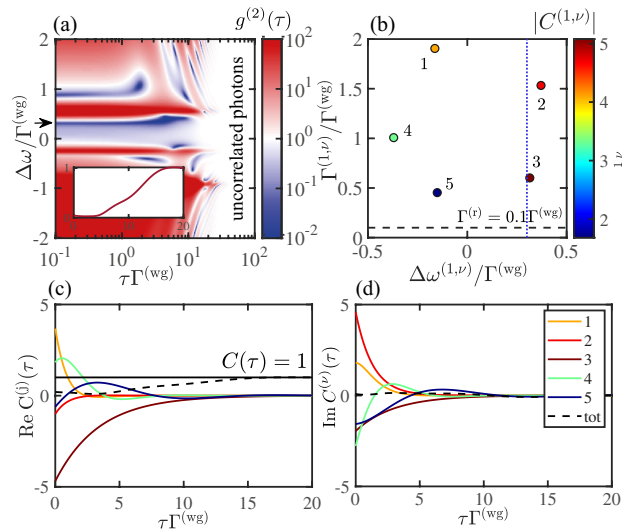}
\caption{(a) Second-order correlation function $g^{(2)}$ versus the photon detuning $\Delta\omega$ and delay time $\tau$, the inset shows $g^{(2)}(\tau)$ for $\Delta\omega=0.3\Gamma^{(\text{wg})}$ specified by an arrow.  (b) Single excitation spectrum of the effective Hamiltonian. Color encodes the magnitude of the corresponding constants $|C^{(1,\nu)}|$. Horizontal black dashed line indicates the radiation losses out of the waveguide $\Gamma^{(\text{r})}$, while a vertical blue dotted line specifies the detunings of the two photons $\Delta\omega=0.3\Gamma^{(\text{wg})}$. (c) Normalized two-photon scattering amplitudes: single-atom non-linearity (1 at. nlin.), contribution from double-excited state of $H^{\text{eff}}$ (2 ex. st.). Real (d) and imaginary (f) parts of the two-photon scattering amplitudes versus delay time $\tau$, color is the same as in $(b)$.  Parameters are: number of atoms $N = 5$, atom-atom distance $a = 0.22 \lambda_{\text{wg}}$, asymmetry parameter $\xi=\Gamma^{(\text{b})}/\Gamma^{(\text{f})}=0.01$, radiation losses $\Gamma^{(\text{r})}=0.1\Gamma^{(\text{wg})}$.
%\commentANP{could we rename ``radiation losses'' e.g. to ``free-space emission losses''? I am used to calling them non-radiative losses in a sense that they do not contribute to the waveguide emission. So the term ``radiation losses'' is a bit confusing to me.}
}
\label{F3}
\end{figure}
%%%%%%%%%%%%%%%%%%%%%%%%%%%%%%%%%%%%%%%%%%%%%
The photon mode propagating in a waveguide is characterized by an intensity distribution across the waveguide \cite{KienOC2004}, polarization, and direction of propagation. The latter, unlike the free space case, is a discrete variable (forward, and backward directions) as the photons are restricted in one dimension. This makes the whole scheme look more realistic for an experimental realization, see the recent review~\cite{sheremet2021waveguide}. In this case, single-excitation effective  Hamiltonian reads as \cite{PichlerPRA2015, FedorovichPRA2022}:
\begin{gather}\label{eq:Heffwqed}
     H_{i,j}^{(\text{eff})} = -i  \begin{cases}
     \Gamma^{(\text{f})} e^{i \phi_{i,j} } & \text{ if $i > j$}, \\
     \Gamma^{(\text{b})} e^{i \phi_{i,j} } & \text{ if $i < j$}, \\
     \dfrac{ \Gamma^{(\text{wg})} + \Gamma^{(\text{r})} }{2}, & \text{ if $i = j$}
    \end{cases},
\end{gather}
where $\phi_{i,j} = k_z |z_i - z_j|$ is the propagation phase, $\Gamma^{(\text{f})} (\Gamma^{(\text{b})})$ is forward (backward) emission rate, $\Gamma^{(\text{wg})}=\Gamma^{(\text{f})}+ \Gamma^{(\text{b})}$ is the emission rate into the waveguide in both directions, $\Gamma^{(\text{r})}$ is the radiation losses out of the waveguide~\cite{KienPRA2005}.
The latter is always present in the 
WQED systems as there is always a chance that an excited atom will emit a photon in free space rather than into the  waveguide. For the state-of-the-art structures with natural atoms coupled to a waveguide such losses are on the order of $\Gamma^{(\text{r})}\sim (10\div 100)\Gamma^{(\text{wg})}$, while in the superconducting qubit platform or even quantum dots coupled to photonic crystal waveguides one may have  $\Gamma^{(\text{r})} \ll \Gamma^{(\text{wg})}$~\cite{sheremet2021waveguide}.
%\commentANP{Danil, could you present the version of Eq.~\eqref{eq:Heff} specifically adapted for our system? That is (1) with 1D Green function (2) with nonreciprocity (3) with also extra decay term $\Gamma^{(\text{r})}$ at the diagonal. I understand that this is kind of obvious, but probably it is easier for readers if the work is self-consistent}.
As is evident from Eq.~\eqref{g2simple}, the presence of large radiative losses leads to an increase of the decay rate of all the single-excitation modes, which makes persistent antibunching impossible (we do not consider here long-period atomic arrays where some modes can be immune to $\Gamma^{(\text{r})}$ due to the Borrmann effect~\cite{Poshakinskiy2021Borrmann}). In this work we consider the case of low radiation losses $\Gamma^{(\text{r})} \ll \Gamma^{(\text{f})}, \Gamma^{(\text{b})}$. There also exist other approaches to achieve  anticorrelations that specifically rely on the opposite regime of high losses $\Gamma^{(\text{r})} \gg \Gamma^{(\text{wg})}$ \cite{MahmoodianPRL2018, prasad2020correlating}.

Here, as in the previous subsection, we will focus on the reflection geometry, as shown in Fig.~ \ref{F1}(b). We consider  $N=5$ atoms equally spaced with the separation of $a = 0.22 \lambda_{\text{wg}}$ ($\lambda_{\text{wg}}$ is the guided mode wavelength) and strong forward emission asymmetry $\Gamma^{(\text{b})}/\Gamma^{(\text{f})}=10^{-2}$. 
Figure~\ref{F3}(a) shows the temporal dependence of the correlation function for backscattered photons depending on the detuning. 
Despite the fact that here we take the system that predominantly emits into the forward direction,  our calculation shows that the backward scattered photons can demonstrate a desired antibunching. Specifically, for the detuning $\Delta\omega=0.3\Gamma^{(\text{wg})}$ (black arrow), the correlation function remains to be $g^{(2)} < 0.1$ for a time $0 \le \tau \Gamma^{(\text{g})} \le 5$, and afterwards reaches the level $0.5$ at approximately $\tau \Gamma^{(\text{g})} \approx 10$, as can be seen in the inset. Figure~\ref{F3} (b) demonstrates that in contrast to the free-space case [Fig.~\ref{F2}(d)], all five single excitation eigenstates contribute significantly to the signal. Moreover, none of these eigenstates are strongly subradiant --- the smallest emission rate is equal to $\Gamma^{(1,\nu=5)} \approx 0.45 \Gamma^{(\text{wg})}$, which is noticeably larger than the radiation losses $\Gamma^{(\text{r})}$. A further inspection of how the contributions from single excitation eigenstates evolve with $\tau$ [Fig. \ref{F3} (c), (d)] demonstrates that long temporal antibunching is a result of a peculiar interference of all of the eigenstates, while the overall lifetime of these correlations is fundamentally limited by the value of radiation losses out of the waveguide $\Gamma^{(1, \nu)} \ge \Gamma^{(\text{r})} = 0.1 \Gamma^{(\text{wg})}$. We would like to term this effect as an \textit{accidental persistent antibunching} as it relies on the destructive interference between several scattering channels.

In partial similarity to the case covered in the previous subsection, here the frequency window of the effect depends on the spectral position of single excitation states with respect to the frequency of the incident photons. For instance, by comparing Fig. \ref{F3} (a), and (b), one can immediately see that this window is rather narrow in the blue region due to a close proximity of states $3, 2$, and much wider in the red region as states $1, 4, 5$ are more spectrally distant.

%{\color{red} DK: point to similarity to accidental BIC?}. \commentANP{Sure! Later on we might consider mentioning this similarity in the introduction, but for now let us add such sentence with a reference here}

\section{Conclusions}\label{sec:conclusions}
We have developed a general analytical expression for a  scattering matrix, characterizing two-photon scattering  on an ensemble of two-level atoms. The approach is based on the electromagnetic Green's function and is suitable for an arbitrary electromagnetic environment. 
In particular, we apply it to atomic arrays in free space and atoms asymmetrically coupled to a waveguide mode.
Next, we have analyzed the photon-photon correlations and suggested two mechanisms to achieve photon antibunching that is persistent on large delay times. The first mechanism is based on the resonance between the incident photons and the single-excitation collective subradiant states of the system, when only a single long-lifetime eigenstate contributes significantly to the scattering amplitude. The second mechanism relies on a destructive interference between the contributions of at least several collective states, none of which are required to be strongly subradiant. We  provide pictorial examples for both mechanisms.  The first mechanism is illustrated for a  free-space array, where we have tuned the frequency and the polarization of the incident photons to selectively excite a single subradiant state. The second mechanism is illustrated for an array of atoms coupled to a waveguide mode in the strongly asymmetric (almost unidirectional) regime.
Both proposed mechanisms are rather general and will hopefully provide useful insights for the rapidly developing quantum optics of atomic arrays.

An interesting possible direction for the future research could be generalization of our results to a strongly driven or a many-photon regime. The lifetime of the  correlations in the interacting quantum systems is an open fundamental problem. Such  phenomena as formation of periodic time-crystalline order \cite{Sacha2017,Mi2021}, accelerated decay to the stationary state \cite{Carollo2021}  and the opposite effect  \cite{Lu2017}, are now actively studied. They involve many-body dynamics that is notoriously hard to analyze.  One of the useful approximations is describing the correlation decay by keeping just a couple of eigenstates of the evolution operator with the smallest decay rates~\cite{Teza2022}. The system considered in our work, where multiple relatively fast decaying eigenstates provide substantial contributions to the correlation dynamics, presents an instructive counterexample to this two-state approximation.

%\section{REFS}
% \cite{PhysRevA.94.053840, lubatsch2009optically, PhysRevA.96.053804, PhysRevA.102.043525, PhysRevB.101.205303, PhysRevA.95.033818, PhysRevB.103.224520, PhysRevA.93.063830, PhysRevLett.121.143601, PhysRevB.85.195463, PhysRevX.7.031024, PhysRevResearch.2.013173, PhysRevB.94.144206, https://doi.org/10.1002/andp.20095211210, okuma2022non, VanTiggelen1999, IVCHENKO199417}

\section{Acknowledgements}

The analytical and numerical calculations done by D.K. have been supported by RSF (project No. 21-72-10107), and by the Priority 2030 Federal Academic Leadership Program.

%{\color{red} Acknowledgements. DK: Most likely, RFBR + Priority 2030.} 

\appendix

\section{Alternative representation of the two-photon kernel $Q$ for general Hamiltonians}

\label{App B}

Relations Eq. \eqref{Sig1exc}, \eqref{Qdef} for the operator $Q(\Omega)$ are sufficient for the efficient calculations, however there exists an alternative way to represent it as a sum over the  two-excitation eigenstates. An expansion of this kind has been done Ref.~\cite{ke2019inelastic} in the context of qubits symmetrically interacting with each other. Here, we present a more general expression that is exactly valid for the most general Hamiltonian, including non-reciprocal ones:
\begin{gather}
    Q(\Omega) = i \left( \Omega + i \Gamma \right) \mathbf{I}  - 4 i \sum_\mu \dfrac{\mathbf{d}^{(R,\mu)} \otimes \mathbf{d}^{(L,\mu)} }{\Omega - E^{(2,\mu)}}, \nonumber\\
    d^{(R,\mu)}_i = \sum\limits_{j} H_{i,j}^{(\text{eff})} \Psi^{(R,\mu)}_{j,i}, \quad d^{(L,\mu)}_i = \sum\limits_{j} \Psi^{(L,\mu)}_{i,j} H_{j,i}^{(\text{eff})},
    \label{Qmatrix2excexp}
\end{gather}
where $\otimes$ is a tensor (outer) product, $\Psi^{(R,\mu)}$ ($\Psi^{(R,\mu)}$) is the two-excitation eigenstate $\nu$ in a two-dimensional representation, with the corresponding complex energy $E^{(2,\mu)}$, and normalized such that $\sum_{i,j}|\Psi^{(R,\mu)}_{i,j}|^2 = 1$. The first term in the expansion above has a meaning of a single atom nonlinearity, while the second term is responsible for a \textit{collective} nonlinearity of an ensemble as it explicitly contains information about the two-excitation sector of $H^{(\text{eff})}$. One can notice that by matching the total energy of two photons $\Omega$ to some doubly excited state $\text{Re}\: E^{(2, \mu)}$ it is possible to significantly enhance the collective non-linearity of the system or even observe doubly excited states in the spectrum \cite{ke2019inelastic}.

We want to note that even though the expansion above is useful for physical understanding, it is not efficient to compute numerically $Q(\Omega)$ using it, and it is much more convenient to use Eq. \eqref{Sig1exc} instead, especially for large atomic ensembles.

\section{Two-photon scattering in free space}
\label{App A}
We start with the definition of a single-photon $S$-matrix:
\begin{gather}
    S^{(1)}_{\lambda', \lambda} = \delta_{\lambda', \lambda} - 2 \pi i T^{(1)}_{\lambda', \lambda}(\omega_{\lambda}) \delta(\omega_{\lambda'} - \omega_{\lambda}),
\end{gather}
where $ T^{(1)}(\omega_{\lambda}) =  V \left( \omega_{\lambda} -  H^{(\text{eff})} \right)^{-1} \hat V$ is a single photon $T$-matrix. The matrix elements are calculated on a state where a single photon in mode $\lambda$ is present, while all $N$ atoms are de-excited $|g\rangle^{\otimes N}$. Note that when the free space case is considered, then $\delta_{\lambda', \lambda} = \dfrac{\left( 2 \pi \right)^3}{V} \delta(\omega' - \omega) \delta(\mathbf{n'} - \mathbf{n}) \delta_{\sigma', \sigma}$, and the singularity formally appears when the forward scattering is considered $\mathbf{n} = \omega_{n'}$, that is not cancelled out. From now on we will only consider the case, when $\mathbf{n'} \ne \mathbf{n}$.
After evaluating the  matrix element of the $ T^{(1)}(\omega_\lambda)$ matrix  given by:
\begin{gather}
    T^{(1)}_{\lambda', \lambda}(\omega_\lambda) = \sum\limits_{n', n} g^*_{\lambda', n'} G_{n', n}(\omega_{\lambda}) g_{\lambda, n},
\end{gather}
one can readily obtain the linear part of the scattered unnormalized wavefunction in the time domain:
\begin{multline}
    \Psi^{(\text{lin})}(t_1, t_2) = \\
    - 4 \pi^2 \bigg(  T^{(1)}_{\mathbf{k}'_1,\sigma'_1; \mathbf{k}_1,\sigma_1}(\omega_1) T^{(1)}_{\mathbf{k}'_2,\sigma'_2; \mathbf{k}_2,\sigma_2}(\omega_2) e^{-i \omega_1 t_1 - i \omega_2 t_2} + \\
    T^{(1)}_{\mathbf{k}'_2, \sigma'_2; \mathbf{k}_1, \sigma_1}(\omega_1) T^{(1)}_{\mathbf{k}'_1, \sigma'_1; \mathbf{k}_2, \sigma_2}(\omega_2) e^{-i \omega_2 t_1 - i \omega_1 t_2} \bigg)\:.
    \label{AppAPsiLin}
\end{multline}
As has been mentioned  before, this answer is valid when $\mathbf{n}'_1 \ne \mathbf{n}_1, \mathbf{n}_2$, $\mathbf{n}'_2 \ne \mathbf{n}_1, \mathbf{n}_2$. From this expression it is already obvious that if one allows the two incident photons to have the different frequencies $\omega_1 \ne  \omega_2$, then, in principle, at certain time difference $|t_1 - t_2|$ this wavefunction might become zero, $\Psi^{(\text{lin})}(t_1, t_2) = 0$. Even though it is generally not a problem and appears due to an obvious destructive interference, when calculating the normalized correlation function $g^{(2)}(t_1, t_2)$ it might lead to somewhat artificial divergencies that appear not due to quantum mechanical bunching, but rather due to a linear term being zero. In order to avoid this, from now on we consider $\omega_1 = \omega_2$.

For the nonlinear part we can write:
\begin{gather}
    \Psi^{(\text{nlin})}(t_1, t_2) = \nonumber\\
    \quad \quad + \int \int 4 \pi M_{\sigma'_1, \sigma'_2; \sigma_1, \sigma_2} \delta(\Omega' - \Omega) e^{- i \omega'_1 t_1 - i \omega'_2 t_2} d\omega'_1 d\omega'_2, \nonumber\\
    M_{\sigma'_1, \sigma'_2; \sigma_1, \sigma_2} = \nonumber\\
    s^+_{\sigma'_1, i}(\omega'_1) s^+_{\sigma'_2, i}(\omega'_2) Q_{i; j}(\Omega) s^-_{\sigma'_1,j}(\omega_1) s^-_{\sigma_2, j}(\omega_2),
    \label{AppAPsinlin}
\end{gather}
where for the matrix elements of $s^-(\omega_1)$ introduced in the main text, we can define:
\begin{gather}
    g_{\mathbf{k}_1, \sigma_1; j} = -i \sqrt{\dfrac{2 \pi \hbar \omega_k}{V}}  \left( \mathbf{d}^{eg}_j \cdot \mathbf{\epsilon}_{\mathbf{k_1}, \sigma_1} \right) e^{i \mathbf{k}_1 \cdot \mathbf{r}_j}\:,
\end{gather}
which is the coupling constant of an atom $j$ with the plane wave having a wavevector $\mathbf{k_1}$, and polarization $\mathbf{\epsilon}_{\mathbf{k_1}, \sigma_1}$. 

Now we need to evaluate the double frequency integral in Eq.~\ref{AppAPsinlin}. One can notice that the frequencies of the outgoing photons $\omega'_1, \omega'_2$ only enter into the operators responsible for the photon emission $s^+_{\sigma'_1, i}(\omega'_1),s^+_{\sigma'_2, i}(\omega'_2)$ and the delta function $\delta(\Omega-\Omega')$. These operators, as Eq. \ref{spsmdefs} from the main text suggests, depend on the single excitation Green's function that has poles at complex-valued eigenfrequencies/eigenenergies $E^{(1, \nu')}$ of the collective single excitation states of the ensemble, namely, \[G_{i,j}(\omega') = \sum\limits_{\nu'=1}^{N} \dfrac{v^{(R,\nu'_1)}_{i} v^{(L,\nu'_1)}_{j}}{\omega' - E^{(1,\nu')}} =  \sum\limits_{\nu'=1}^{N} \dfrac{g^{(\nu')}_{i,j}}{\omega' - E^{(1,\nu')}}\:.\] Using it, we can carry out the double integral:
\begin{widetext}
\begin{gather}
    \int \int R_i(\omega'_1, \omega'_2) s^+_{\sigma'_1, i}(\omega'_1) s^+_{\sigma'_2, i}(\omega'_2) \delta(\omega'_1+\omega'_2-\omega_1-\omega_2)  e^{- i \omega'_1 t_1 - i \omega'_2 t_2} d \omega'_1 d \omega'_2 = \nonumber\\
    \int R_{i}(\omega'_1, \Omega-\omega'_1) \sum\limits_{i'_1,i'_2} g_{\mathbf{k}'_1,\sigma'_1;i'_1}^* g_{\mathbf{k}'_2,\sigma'_2;i'_2}^* \sum\limits_{\nu'_1, \nu'_2} \dfrac{ g^{(\nu'_1)}_{i'_1, i} }{\omega'_1 - E^{(1, \nu'_1)}} \dfrac{ g^{(\nu'_2)}_{i'_2, i} }{\Omega-\omega'_1 - E^{(1, \nu'_2)}} e^{- i \Omega t'_2 - i \omega'_1 (t'_1 - t'_2)} d\omega'_1,
\end{gather}
\end{widetext}
where $R_{i}(\omega'_1, \omega'_2)$ is the remaining part of the Eq. \ref{AppAPsinlin}. The remaining integral over $\omega'_1$ can be straightforwardly evaluated  using the residue theorem, and on which half-plane the contour should be enclosed depends on the sign of $t'_1-t'_2$. For $t'_1>t'_2$ we choose the lower one ($\text{Im}\: \omega'_1<0$), which contains poles at $\omega'_1 = E^{(1, \nu'_1)}$, while in the opposite case the upper plane is chosen with poles at $\omega'_1 = -\Omega + E^{(1, \nu'_2)}$. Finally, we arrive at the following result:
\begin{widetext}
    \begin{multline}
    \Psi^{(\text{nlin})}(t_1, t_2) \approx - 8i \pi^2 \sum\limits_{i'_1,i'_2} g^*_{\mathbf{n}'_1, \sigma'_1; i'_1} g^*_{\mathbf{n}'_2, \sigma'_2; i'_2} \left(  \sum\limits_{\nu'_1, \nu'_2} \dfrac{g^{(\nu'_1)}_{i'_1, i} g^{(\nu'_2)}_{i'_2, i} e^{- i E^{(1, \nu'_1/\nu'_2)}(t_> - t_<)} e^{-i \Omega t_< } }{\Omega - E^{(1, \nu'_1)} - E^{(1, \nu'_2)}} \right) Q_{i, j}(\Omega) \cdot \\
    \left( \sum\limits_{i_1} G_{j, i_1}(\omega_1) g_{ \mathbf{n}_1, \sigma_1; i_1}  \right) \left( \sum\limits_{i_2} G_{j, i_2}(\omega_2) g_{\mathbf{n}_2, \sigma_2; i_2} \right),
    \end{multline}
\end{widetext}
where $t_>(t_<)$ is the largest (smallest) of the detection times $t_1, t_2$. During the derivation we have also used a near-resonant approximation ($|\omega_{j} - \omega_0| \ll \omega_0$), which formally means that one has to make the replacement $\omega_{j} \to \omega_0$, and $\mathbf{k}_j \to k_0 \mathbf{n}_j$ in the definitions of coupling constants. 

We would like to make one more comment. The notation $E^{(1, \nu'_1/\nu'_2)}$ means that one has to pick $\nu'_1$ when $t'_1>t'_2$, and $\nu'_2$ otherwise. This is important because if one wants to derive a formula similar to Eq. $\ref{g2simple}$, then the indices $\nu'_1, \nu'_2$ get coupled to the quantum numbers of the outgoing photons $(\mathbf{k}'_1,\sigma'_1), (\mathbf{k}'_2,\sigma'_2)$, and this alternation $\nu'_1/\nu'_2$ is required to keep the photon wavefunction symmetric with respect to the total exchange of quantum numbers of two photons $1 \longleftrightarrow 2$. However, if the two outgoing photons are totally identical, then one can put either $\nu'_1$ or $\nu'_2$ in the above expression.

Since the general expression is rather cumbersome, we will consider the case of twin-photon scattering, meaning that $\omega_1 = \omega_2$, $\mathbf{n}_1'=\mathbf{n}_2'$, $\mathbf{n}_1=\mathbf{n}_2$. The normalized second-order correlation function is then given by:
\begin{multline}
    g^{(2)}(\tau) = \left|1 +  \frac{\Psi^{(\text{nlin})}(t_1,t_2)}{\Psi^{(\text{lin})}(t_1, t_2)} \right|^2 = \nonumber\\
    \left| 1 - \sum\limits_{\nu} C^{(\nu)}_{\mathbf{n}', \sigma'; \mathbf{n}, \sigma} (\omega) e^{-i \left( E^{(1, \nu)} - \omega \right) \tau } \right|^2\:,
\end{multline}
%{\color{red} Not sure I need to provide explicit def for $C^{\nu}$}.
where the explicit expression for $C^{(\nu)}_{\mathbf{n}', \sigma'; \mathbf{n}, \sigma} (\omega)$ has the form:
\begin{widetext}
    \begin{multline}
        C^{(\nu'_1)}_{\mathbf{n}', \sigma'; \mathbf{n}, \sigma} (\omega) = \\
        \dfrac{ \big( \mathbf{\epsilon}_{\mathbf{k'}, \sigma'}^* \cdot \mathbf{d}^{ge}_{i'_1} \big) e^{- i k_0 \mathbf{n}' \cdot \mathbf{r}_{i'_1} } \big( \mathbf{\epsilon}_{\mathbf{n'}, \sigma'}^* \cdot \mathbf{d}^{ge}_{i'_2} \big) e^{- i k_0 \mathbf{n}' \cdot \mathbf{r}_{i'_2} } \bigg( \sum\limits_{\nu'_2} \dfrac{ g^{(\nu'_1)}_{i'_1, i} g^{(\nu'_2)}_{i'_2, i} }{\Omega - E^{(1, \nu'_1)} - E^{(1, \nu'_2)}} \bigg) Q_{i,j}(\Omega) \left( \sum\limits_{i_1} G_{j,i_1}(\omega) \left( \mathbf{d}^{eg}_{i_1} \cdot 
     \mathbf{\epsilon}_{\mathbf{n}, \sigma} \right) e^{i k_0 \mathbf{n} \cdot \mathbf{r}_{i_1}}  \right)^2 }{\left( \big( \mathbf{\epsilon}_{\mathbf{k'}, \sigma'}^* \cdot \mathbf{d}^{ge}_{m'} \big) e^{- i k_0 \mathbf{n}' \cdot \mathbf{r}_{m'} } G_{m', m}(\omega) \big( \mathbf{d}^{eg}_{m} \cdot 
     \mathbf{\epsilon}_{\mathbf{n}, \sigma} \big) e^{i k_0 \mathbf{n} \cdot \mathbf{r}_m}  \right)^2 }.
     \label{Cconst}
    \end{multline}
\end{widetext}

In order to double check the obtained formula, one can compute $g^{(2)}(\tau)$ for a single atom in free space, for which there is only one single-excited eigenstate: $E^{(1,1)} = \omega_0 - i \Gamma/2$, $g^{(1)}_{1,1} = 1$, $G(\omega) = (\omega - \omega_0 + i \Gamma/2 )^{-1}$. By substituting this into the expression above for $C^{(\nu)}_{\mathbf{n}', \sigma'; \mathbf{n}, \sigma} (\omega)$, one can obtain $g^{(2)}(\tau) = \big| 1 - e^{ -i(\omega_0 - \omega)\tau - \frac{\Gamma}{2} \tau } \big|^2$, which is precisely what we would expect.

The obtained result can be easily modified in order to be suitable to use in WQED systems. One only needs to consider that $g_{\lambda,i} = \sqrt{\Gamma_{\sigma,i}}$, where $\Gamma_{\sigma,i}$ is the emission rate of atom $i$ into the mode propagating in direction $\sigma$. Here we assume that the waveguide is a single mode one (fixed polarization and distribution of the fields), such that the only quantum number for photons is the direction of propagation $\sigma$, but this is straightforward to generalize.

\bibliography{Biblio}% Produces the bibliography via BibTeX.

\end{document}